\documentclass[conference]{IEEEtran}
\IEEEoverridecommandlockouts
\usepackage{cite}
\usepackage{amsmath,amssymb,amsfonts,dsfont}
\usepackage{algorithmic}
\usepackage{graphicx}
\usepackage{textcomp}
\usepackage{xcolor}
\usepackage{subfigure,dblfloatfix}

\newcommand{\Fig}[1]{Figure~\ref{fig:#1}}

\newcommand{\Sec}[1]{Sec.~\ref{sec:#1}}

\newcommand{\Eq}[1]{(\ref{eq:#1})}

\newcommand{\ind}[1]{\mathds{1}_{[#1]}}

\input{my_math.sty}
\newcommand{\Mbs}{M_{\rm BS}}
\newcommand{\Mue}{M_{\rm UE}}
\newcommand{\Mmn}{M_{\rm MN}}

\def\BibTeX{{\rm B\kern-.05em{\sc i\kern-.025em b}\kern-.08em
    T\kern-.1667em\lower.7ex\hbox{E}\kern-.125emX}}

\begin{document}

\title{
Eavesdropping with Intelligent Reflective Surfaces:\\Threats and Defense Strategies
}

\author{\IEEEauthorblockN{Francesco Malandrino}
\IEEEauthorblockA{CNR-IEIIT \\
Torino, Italy}
\and
\IEEEauthorblockN{Alessandro Nordio}
\IEEEauthorblockA{CNR-IEIIT \\
Torino, Italy}
\and
\IEEEauthorblockN{Carla Fabiana Chiasserini}
\IEEEauthorblockA{Politecnico di Torino and CNR-IEIIT \\
Torino, Italy}
}

\maketitle

\begin{abstract}
Intelligent reflecting
surfaces (IRSs) have several prominent advantages, including improving 
the level of wireless communications security and privacy. In this work, 
we focus on this aspect and envision a strategy to counteract the presence of passive eavesdroppers 
overhearing transmissions from a base station towards legitimate users.
Unlike most of the existing works addressing passive eavesdropping, 
the strategy we consider has low complexity and is suitable for scenarios where nodes are 
equipped with a limited number of antennas. Through our performance evaluation, we highlight the trade-off
between  the legitimate users' data rate and 
secrecy rate, and how the system parameters affect such a trade-off.
\end{abstract}


\section{Introduction\label{sec:intro}}
It is expected that future generation of mobile communications (6G)
will exploit THz frequencies (e.g., 
0.1–10~THz~\cite{Aluoini-mmw-thz, Akyildiz-THz}) for indoor as well as
outdoor applications.  THz communications can indeed offer 
very high data rates, although over short 
distances, due to harsh propagation conditions and severe path loss.
To circumvent these problems, massive MIMO (M-MIMO) communication and
beamforming techniques can be exploited to concentrate the transmitted
power towards the intended receiver. Further, the use of intelligent reflecting
surfaces (IRSs)~\cite{Wu} has emerged as a way to enable smart radio environments (SRE)~\cite{DiRenzo}, i.e., 
to control and adapt the radio environment to the communication between a transmitter and a receiver, 
aiming at optimizing the performance.  

IRSs are passive beamforming devices, composed of a large number of
low-cost antennas, that receive signals from sources, customize them
by basic operations, and then forward it toward desired
directions~\cite{Liaskos,Liaskos-3,Alsharif2020}.
As discussed in~\cite{zhou2020enabling}, IRSs can be efficiently used
to improve the security and privacy of wireless communications,
as they can make the channel better for legitimate users,
and worse for malicious ones.

As an example, the authors of~\cite{zhou2020user} target the case of {\em aligned} eavesdroppers, 
lying between the transmitter and the legitimate receiver: in this case, the authors 
envision avoiding direct transmissions, and using IRSs to maximize the secrecy rate. 
Jamming is an effective, even if harsh, method to improve privacy by making the eavesdropper's 
channel worse: as an example, \cite{guan2020intelligent}~envisions using IRSs to both serve 
legitimate users and jam the malicious one, maximizing the secrecy rate subject to power constraints.
In MIMO scenarios, passive eavesdroppers can be blinded through standard beamforming techniques, 
thanks to the so-called secrecy-for-free property of MIMO systems with large antenna arrays. 
Several recent works, including~\cite{bereyhi2020secure}, aim at achieving the same security level 
against active attackers, by leveraging filtering techniques and the fact that legitimate 
and malicious node are statistically distinguishable from each other. 
In a similar scenario, \cite{dong2020secure} presents an alternating optimization that jointly 
optimizes both transmitter and IRS parameters in order to maximize the secrecy rate.

In this work, we investigate the secrecy performance of IRS-based communications, considering 
the presence of a malicious receiver passively overhearing the downlink transmission intended 
for a legitimate user. 
Unlike previous works, we consider a low-complexity scheme to make the system robust to eavesdropping, which 
leverages the key observation that, in virtually all real-world scenarios, 
IRSs point {\em towards a UE}. It follows that we can study the way IRSs are {\em used} 
to steer the signal, 
as opposed to the standard approach of optimizing the complex coefficients of the 
{\em matrices} describing the behavior of terminals and IRSs. The result is a very significant 
reduction of the solution space to explore, hence, of the complexity of the decision-making 
process as a whole.

The rest of the paper is organized as follows. 
\Sec{model} introduces the model of the system under study and behavior of the malicious node.
\Sec{problem} presents the problem formulation and the configuration-switch strategy adopted in the system, 
while \Sec{peva} shows the relevant tradeoffs.
Finally, \Sec{concl} concludes the paper.

\begin{figure} %
\includegraphics[width=0.45\textwidth]{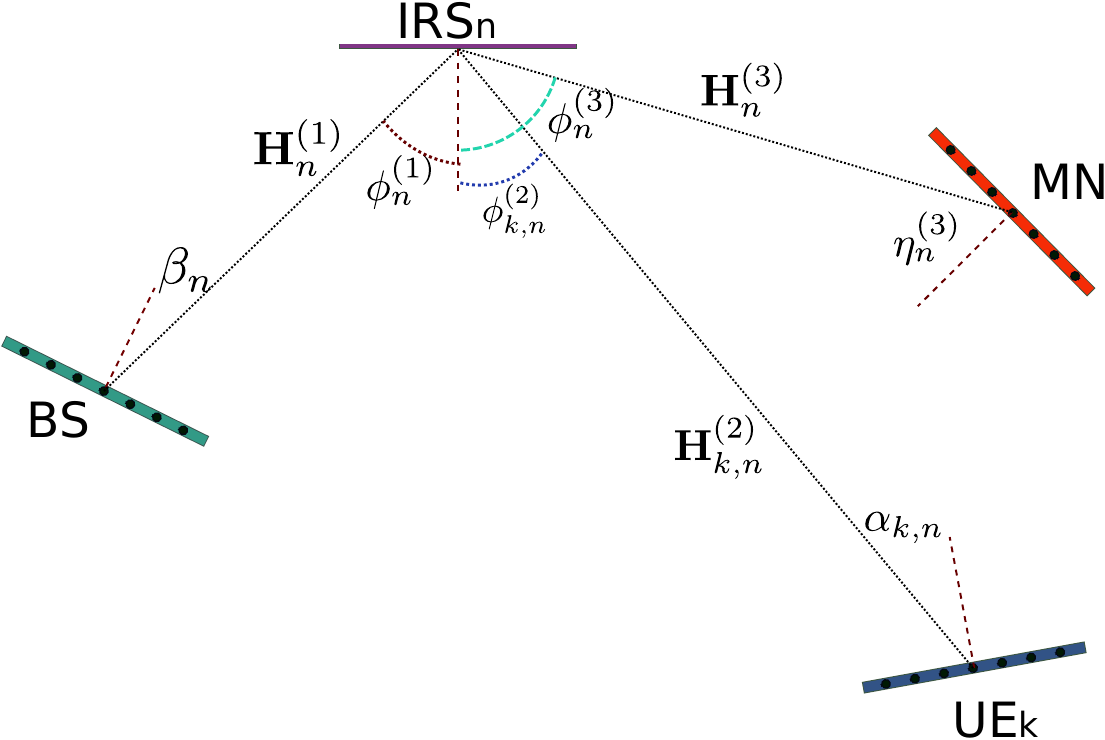}
\centering
\caption{\label{fig:model} Communication model: a base station (BS) is transmitting toward a user equipment (UE$_k$), 
thanks to the help of the IRSs. UE$_k$ is the victim of the malicious node (MN), which intercepts the signals 
reflected by the IRSs.}
\end{figure}

\section{System Model}
\label{sec:model}

We consider a wireless network operating in the THz bands, composed of: 
\begin{itemize}
\item a base station (BS) whose uniform linear array (ULA) of antennas has $\Mbs$~isotropic
  elements and transmitting $K$ streams, one for each legitimate
  user;
\item $K$ legitimate users (UE), each equipped with an ULA composed of $\Mue$
  isotropic antenna elements;
\item $N$ IRSs ($N\ge K$); the $n$-th IRS includes $L_n^2$ antenna elements (or
  meta-atoms), arranged in a $L_n\times L_n$ square grid or uniform
  planar array (UPA). The IRSs contribute to the BS-UEs communication
  by appropriately forwarding the BS signal toward the users. 
  Notice that, given $K$ legitimate UEs, at any time instant only $K$ IRSs are used.
\item a passive eavesdropper (or malicious node, MN), whose ULA is composed of
  $\Mmn$ isotropic antenna elements. The goal of the MN is to eavesdrop the communication from 
  the BS towards one of the $K$ legitimate UE, by intercepting the signals reflected by the IRSs. 
To do so, the MN exploits the directivity provided by its ULA by pointing
  it towards the IRS serving the UE that the MN wants to eavesdrop. 
\end{itemize}

In the remainder of this section, we characterize the main elements of the system we consider, 
namely, the channel and IRSs (\Sec{sub-channel}), the nodes and their behavior (\Sec{sub-nodes}), 
the metrics we consider (\Sec{sub-metrics}), and the time dimension (\Sec{sub-time}).

\subsection{Channel and IRS}
\label{sec:sub-channel}

The elements of the ULAs and of the IRSs are assumed to be separated
by $\lambda/2$ where $\lambda$ is the wavelength of the signal
carrier.  We assume that no line-of-sight (LoS) path exists between the
BS and the UEs. However, communication is made possible by the ability
of the IRSs to reflect the BS signal towards the
users~\cite{noi-icc20}.  All IRSs and user nodes,
including the MN, are assumed to have the same height above ground, as depicted in
Figure~\ref{fig:model}. This assumption allows simplifying the
discussion and the notation while capturing the key aspects of the
system. In the following, we detail the communication chain
as well as the channel model.

\noindent {\bf Communication channel.} While in many works dealing with
GHz communications, the channel matrix connecting two
multi-antenna devices is often modeled according to Rayleigh or Rice
distributions, in the THz bands the channel statistic is not yet
completely characterized.  Moreover, at such high frequencies the
signal suffers from strong free-space attenuation, and is blocked even
by small solid obstacles.  In practice, the receiver needs to be in line-of-sight (LoS)
with the transmitter to be able to communicate.  Also, recent
studies~\cite{Xing} highlight that already at sub-THz frequencies all scattered
components and multipath effects can be neglected.  This conclusion is
also supported by the fact that, typically, both the transmitter and
the receiver employ massive beamforming techniques in order to
concentrate the signal energy along a specific direction and compensate 
for high path losses. In such conditions, the channel matrix
between any two devices can thus be modeled as: 
\begin{equation}
  \label{eq:channel_model}
  \Hm^{\rm LOS} = a c \pv \qv\Herm
\end{equation}
where the scalar $a$ takes into account large scale fading effects due
to, e.g., obstacles temporarily crossing the LoS path between
transmitter and receiver. The coefficient $c$ instead 
accounts for the attenuation and phase rotation due to propagation.
More specifically, let $d$ be the distance between the transmitting
and the receiving device and $G$ be the array gain of one of them.
Then the expression for $c$ is given by:
    \begin{equation}\label{eq:c_Area}
      c = \sqrt{\frac{G A}{4\pi d^2}}\ee^{\jj \frac{2 \pi}{\lambda}d }
    \end{equation}
where $A$ is the effective area of the other device.
    Finally, $\pv$ and $\qv$ are norm-1 vectors representing,
    respectively, the spatial signatures of the receive and transmit
    antenna arrays.  The spatial signature of a ULA composed of $M_{\rm z}$ ($z=$BS,UE,MN)
    isotropic elements spaced by $\lambda/2$ and observed
    from an angle $\beta$ (measured with respect to a direction
    orthogonal to the ULA), is given by the size-$M_z$ vector
    $\sv(\beta,M_z)$, whose $m$-th element is given by
\begin{equation}
  \label{eq:signature}
  [\sv(\beta,M_z)]_m = \frac{1}{\sqrt{M_z}}\ee^{-\jj\frac{\pi}{2} (M_z-1)\sin \beta}\ee^{-\jj \pi (m-1)\sin \beta}\,.
\end{equation}
This relation applies to devices equipped with ULAs such as the BS,
the UEs and the MN. However, it can also be applied to IRSs since
their planar configuration can be viewed as a superposition of several
ULAs.

\begin{figure} %
\includegraphics[width=0.45\textwidth]{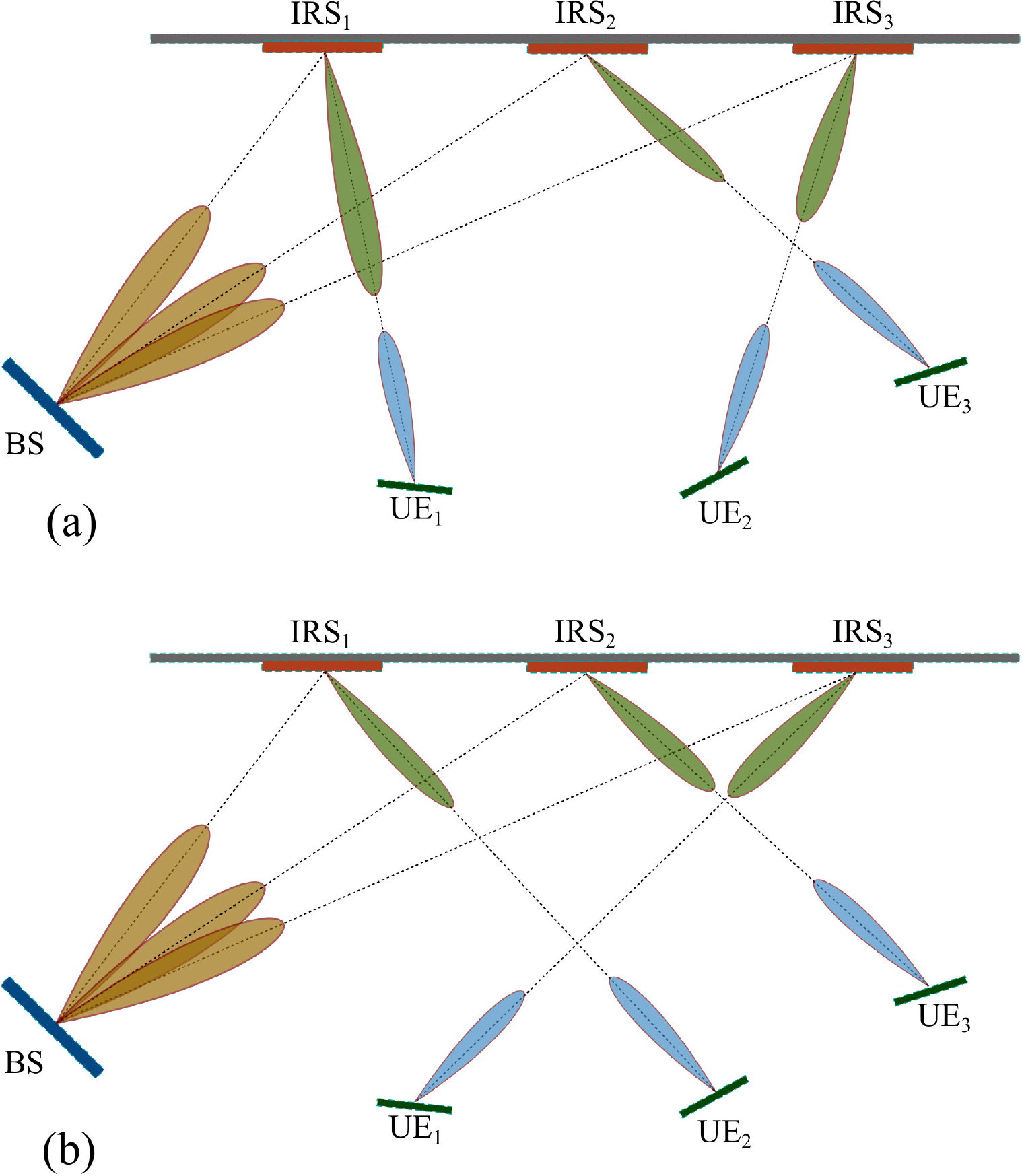}
\centering
\caption{\label{fig:permutations}Two possible permutations for a
  network with $N=K=3$. The configuration in (a) corresponds to the
  map $\nu_a(1)=1$, $\nu_a(2)=3$, and $\nu_a(3)=2$, while the configuration
  in (b) corresponds to the map $\nu_b(1)=3$, $\nu_b(2)=1$, and
  $\nu_b(3)=2$.}
\end{figure}

\noindent{\bf IRS characterization.} IRSs are made of meta-atoms
(modeled as elementary spherical scatterer) whose scattered
electromagnetic field holds in the far-field regime~\cite{direnzo2,Ozdogan}.  The $n$-th IRS, $n=1,\ldots,N$, is composed of $L_n^2$
meta-atoms arranged in an $L_n\times L_n$ square grid.
The area of the $n$-th IRS is, thus, given by
$A_n=L_n^2\lambda^2/4$.  The meta-atom at position $(\ell,\ell')$ in
the $n$-th IRS applies a (controlled) continuous phase shift,
$\theta_{n,\ell,\ell'}$ to the signal impinging on it. In many works
that assume rich scattering communication channels, such phase-shifts
are independently optimized in order to maximize some performance
figures. However, under the channel model in~\eqref{eq:channel_model},
phase-shifts are related to each other~\cite{scattermimo,
  Wu-Zhang} according to the linear equation:
\begin{equation}\label{eq:phase_shift}
  \theta_{n,\ell,\ell'} = \pi q_n\left(\ell-1-\frac{L_n-1}{2}\right) + \psi_n
\end{equation}
where $q_n$ and $\psi_n$ control, respectively, the direction and the
phase of the reflected signal. For simplicity, we arrange the phase
shifts $\theta_{n,\ell,\ell'}$ in the diagonal matrix
\[  \bar{\Thetam}_n = \Id_{L_n}\otimes \Thetam_n \]
where $\Thetam_n=\diag(\theta_{n,1,\ell'},\ldots, \theta_{n,L_n,\ell'})$, $\otimes$ denotes
the Kronecker product, and $\Id_{L_n}$ is the identity matrix of size $L_n$.
As an example, let $\phi^{(1)}_n$ be the
angle of arrival (AoA) of the BS signals on the $n$-th IRS and let
$\phi^{(2)}_{n,k}$ be the direction of the $k$-th IRS as observed from
the $n$-th IRS (see Figure~\ref{fig:model}). Then, in order to let the $n$-IRS reflect
the BS signal towards the $k$-th UE, we set~\cite{noi-icc20}
\begin{equation}
  q_n = \sin \phi^{(1)}_n - \sin \phi^{(2)}_{n,k}\,. \label{eq:q_n}
\end{equation}
In a practical case, a given IRS serves a single UE; thus, we define as 
{\em permutation} a mapping between the set of UEs and the set of IRSs:
\begin{equation} 
\nonumber
\nu_p : \{1,\ldots, K\} \to \{1,\ldots,N\}. 
\end{equation}
Under this map, IRS $\nu_p(k)$ forwards the BS signal
towards the $k$-th UE and, by symmetry, the $k$-th UE points its beam
towards the $\nu_p(k)$-th IRS.  The number
of possible permutations (or maps) is then $P=N!/(N-K)!$ and its set
is denoted by $\Pc$. Moreover, we also denote by $\bar{\Theta}_{n,p}$
the matrix of the $n$-IRS phase-shifts under the permutation 
$p=1,\ldots,P$. An example of possible permutations for a network
composed of $N=3$ IRSs and $K=3$ UEs is depicted in
Figure~\ref{fig:permutations}.

\subsection{Network nodes}
\label{sec:sub-nodes}

\noindent{\bf Base station.}
The BS transmit a signal with bandwidth $B$ and wavelength
$\lambda$. Such signal contains $K$ data streams, one for each UE.
Let~$x_k$ be the zero-mean unit variance Gaussian complex
i.i.d. random symbol generated for the $k$-th stream at a given
time. Also, let $\gammav_k$, be the beamforming vector employed for
transmitting $x_k$.  Then, the signal transmitted by the BS is given
by: %
\begin{equation} 
\label{eq:t}
\tv= \Gammam \xv 
\end{equation}
where $\Gammam = [\gammav_1, \ldots, \gammav_K]$,
$\xv=[x_1,\ldots, x_K]\Tran$. We assume that the total transmit power
is limited by $\EE[|\tv|^2] = \|\Gammam\|_{\rm F}^2 \le P_t$,
where $\|\cdot\|_{\rm F}$ denotes the Frobenius norm.

\noindent{\bf Legitimate receivers (UEs).} The
signal received by the $k$-th UE under the $p$-th map is given by
\begin{eqnarray}
  r_{k,p} &=& \underbrace{\fv_{k,p}\Herm \sum_{n=1}^N\Hm^{(2)}_{k,n}\bar{\Thetam}_{n,p}\Hm^{(1)}_n}_{\widetilde{\hv}_{n,p}\Herm}\tv +n_k\label{eq:rk}
\end{eqnarray}
where $n_k\sim\Nc_\CC(0,N_0 B)$ is additive Gaussian complex noise,
whose power spectral density is $N_0$. $B$ is the signal bandwidth and
$\fv_{k,p}$ is the beamforming vector at the $k$-th UE. According to
the channel model in~\eqref{eq:channel_model},
$\Hm^{(1)}_n = a_n^{(1)}c_n^{(1)}\pv_n^{(1)}{\qv_n^{(1)}}\Herm$ is the
channel connecting the BS to the $n$-th IRS and
$\Hm^{(2)}_{k,n}=a^{(2)}_{k,n}c^{(2)}_{k,n}\pv_{n,k}^{(2)}{\qv_{n,k}^{(2)}}\Herm$
is the channel matrix connecting the $n$-th IRS to the $k$-th UE.  In
particular, we have $\qv_n^{(1)}=\sv(\beta_n,\Mbs)$,
$\pv_n^{(1)}=\frac{1}{\sqrt{L_n}}\onev_{L_n}\otimes
\bar{\pv}_n^{(1)}$, $\bar{\pv}_n^{(1)}=\sv(\phi^{(1)}_n,L_n)$,
$\pv_{k,n}=\sv(\alpha_{k,n},\Mue)$,
$\qv_{n,k}^{(2)}= \frac{1}{\sqrt{L_n}}\onev_{L_n}\otimes
\bar{\qv}_{k,n}^{(2)}$ and
$\bar{\qv}_{k,n}^{(2)}=\sv(\phi_{k,n}^{(2)},L_n)$.  
  In the above expressions, $\otimes$ denotes the Kronecker product
  and $\onev_{L_n}$ is an all-ones column vectors of size $L_n$.
  Notice that, by assuming that all  
  network nodes have the same height over the ground, the $n$-th IRS
  can be viewed as a superposition of $L_n$ identical ULAs. Thus, its
  spatial signature can be written in compact form by using the
  Kronecker product, as indicated above.

The angles $\beta_n$, $\phi_n^{(1)}$,
$\phi_{k,n}^{(2)}$, and $\alpha_{k,n}$ are specified in
Figure~\ref{fig:model}.  Furthermore,
$c^{(1)}_n=\frac{\sqrt{\Mbs A_n}}{\sqrt{4\pi} d^{(1)}_n}\ee^{\jj
  \frac{2 \pi}{\lambda}d^{(1)}_n}$ and
$c^{(2)}_{k,n} = \frac{\sqrt{\Mue A_n}}{\sqrt{4\pi}
  d^{(2)}_{n,k}}\ee^{\jj \frac{2 \pi}{\lambda}d^{(2)}_{n,k}}$ where
$d^{(1)}_n$ and $d^{(2)}_{n,k}$ are, respectively, the distance
between the BS and the $n$-th IRS, and the distance between the $n$-th
IRS and the $k$-th UE.

We assume that the UE ULA is only capable of analog beamforming. Thus,
the norm-1 vector $\fv_{k,p}$ is defined as
$\fv_{k,p}=\sv(\alpha_{k,n},\Mue)$ where $n=\nu_p(k)$,
i.e., the beam generated by the $k$-th UE ULA points to the
 $\nu_p(k)$-th IRS.  By collecting in the vector $\rv_p$
the signals received by the $K$ UEs and by recalling~\eqref{eq:t}, we
can write
\begin{eqnarray}
  \rv_p &=& \widetilde{\Hm}_p\Herm \tv + \nv\non
      &=& \widetilde{\Hm}_p\Herm \Gammam\xv + \nv\label{eq:rv}
\end{eqnarray}
where $\widetilde{\Hm}_p=[\widetilde{\hv}_{1,p},\ldots,\widetilde{\hv}_{K,p}]$ and $\nv = [n_1,\ldots,n_K]\Tran$.

\noindent{\bf Malicious node.}
By eavesdropping the communication, the MN acts as an additional
receiver. When the $p$-th map is applied and the
MN ULA points to the $n$-th IRS, the received signal can be written similarly 
to~\eqref{eq:rk}, as
\begin{eqnarray}
  o_{n,p}  &=& \bv_n\Herm \sum_{m=1}^N \Hm_m^{(3)}\bar{\Thetam}_{m,p}\Hm_m^{(1)}\tv + \zeta \non
     &=& \widetilde{\bv}_{n,p}\Herm \tv + \zeta    
\end{eqnarray}
where $\Hm_n^{(3)}=a_n^{(3)}c_n^{(3)}\pv_n^{(3)}{\qv_n^{(3)}}\Herm$ is
the channel matrix connecting the $n$-th IRS to the MN,
$\pv_n^{(3)}=\sv(\eta_n,\Mmn)$,
$\qv_n^{(3)}=\frac{1}{\sqrt{L_n}}\onev_{L_n}\otimes
\bar{\qv}_n^{(3)}$, $\bar{\qv}_n^{(3)}=\sv(\phi^{(3)}_n,L_n)$ (see
Figure~\ref{fig:model}).  Also,
$c_n^{(3)}=\frac{\sqrt{\Mmn A_n}}{\sqrt{4\pi} d}\ee^{\jj \frac{2
  \pi}{\lambda}d^{(3)}_n}$ where $d^{(3)}_n$ is the distance between
the MN and the $n$-th IRS. Finally,
$\zeta\sim\Nc_\CC(0,N_0B)$ represents the additive noise
at the receiver and $\bv_n=\sv(\eta_n,\Mmn)$ is the norm-1 beamforming vector.

\subsection{Metrics of interest}
\label{sec:sub-metrics}

For each combination of IRS-to-UE assignment, we are interested in deriving two main metrics, 
namely, the SINR (hence, the data rate) and the secrecy rate.

The SINR achieved at each UE 
depends on the precoding strategy employed at the BS, i.e., on the
choice of the precoder $\Gammam$.  For example, the zero-forcing (ZF)
precoder permits to remove the inter-user interference while providing
good (although not optimal) performance. Under the $p$-th map,
 $\nu_p$, the ZF precoder is obtained by solving for $\Gammam_p$ the equation
$\widetilde{\Hm}_p\Herm\Gammam_p=\mu\Qm$ where
$\Qm=\diag(q_1,\ldots, q_K)$ is an arbitrary positive diagonal matrix and the scalar $\mu$
should be set so as to satisfy the power constraint $\|\Gammam_p\|_{\rm F}^2 = P_t$.
Its expression is given by: 
\begin{equation}\label{eq:Gamma}
  \Gammam_p \triangleq \frac{\sqrt{P_t}\widetilde{\Hm}_p^+\Qm^{1/2}}{\|\widetilde{\Hm}_p\Qm^{1/2}\|_{\rm F}}\,.
\end{equation}
where $\widetilde{\Hm}_p^+=\widetilde{\Hm}_p\Herm(\widetilde{\Hm}_p\widetilde{\Hm}_p\Herm)^{-1}$
is the Moore-Penrose pseudo-inverse of $\widetilde{\Hm}_p$.  Then, the
SINR at the $k$-th UE is given by,
\begin{equation}
{\rm SINR}^{{\rm UE}}_{k,p} = \frac{P_t q_k}{N_0B\|\widetilde{\Hm}_p\Qm^{1/2}\|^2_{\rm F}} 
\end{equation}
Similarly, we can write the SINR at the MN when the latter points its ULA to the $n$-th
IRS while eavesdropping the $k$-th data stream, as
\begin{equation}
{\rm SINR}^{{\rm MN}}_{n,k,p} = \frac{q_k|\widetilde{\bv}_{n,p}\Herm \gammav_{k,p}|^2}
{\sum_{h\neq k}q_h|\widetilde{\bv}_{n,p}\Herm \gammav_{h,p}|^2+N_0B}\,.
\end{equation}
where $\gammav_{k,p}$ is the $k$-th column of $\Gammam_p$ whose
expression is given by~\eqref{eq:Gamma}. 
The data rate for UE $k$ under the $p$-map can be computed as 
\begin{equation}
R(k,p)=B\log_2\left(1+{\rm SINR}^{{\rm UE}}_{k,p}\right).
\end{equation}
Finally, the secrecy
rate (SR) obtained when the MN eavesdrops the $k$-th stream by
pointing its antenna to the $n$-th IRS, under the $p$-map, is given by
\begin{eqnarray}
  {\rm SR}(n,k,p) = \max      \big\{ 0, R(k,p) \mathord{-}  B\log_2(1\mathord{+}{\rm SINR}^{{\rm MN}}_{n,k,p})  \big\}
\label{eq:SR}
\end{eqnarray}
The $\max$ operator in~\eqref{eq:SR} is required since, under certain
circumstances, ${\rm SINR}^{{\rm MN}}_{n,k,p}$ might be larger than
${\rm SINR}^{{\rm UE}}_{k,p}$.

\subsection{Permutations and time}
\label{sec:sub-time}

We define a set of IRS-to-UE assignments as a {\em permutation};
intuitively, each permutation corresponds to one way to serve the
users. Given the set~$\Pc$ of {\em all} possible permutations, the BS
chooses a set~$\bar{\Pc}$ of permutations to activate, as well as a
criterion that legitimate nodes shall follow to determine the next
permutation to move to. In other words, legitimate nodes will always
know the next permutation to use (e.g., because they follow the same
hash chain~\cite{hussain2009key}), while the eavesdropper can not. We
further assume that all chosen permutations are used with equal
probability, and that they are notified to legitimate users in a
secure manner, while the eavesdropper has no way of knowing the next
permutation in advance. As noted earlier, hash chains allow to attain
both goals.

Concerning time, we normalize everything to the time it takes to receivers (legitimate or not) 
to switch from one configuration to another, and call that one {\em time unit}. The BS also sets 
the number of time units the legitimate users should stay with each permutation. 
As for the eavesdropper, we consider the most unfavorable scenario for the legitimate users and assume that 
MN has already estimated the probability with which its victim is served by each IRS, and that it can 
leverage such information.

\section{Problem Formulation and Solution Concept}
\label{sec:problem}

In this section, we formulate the problem of optimizing the decisions made by 
the BS,
i.e., the choice of the permutations to use and the time to wait
before switching to the next permutation, in order to obtain the best
possible secrecy rate {\em and} the required data rate.

Given the set~$\Pc$ of permutations, we can indicate
with~$\nu_p(k)\in[1\dots N]$ the IRS used to serve user~$k$ under
permutation~$p$.  We also know the rate~$R(k,p)$, i.e., the rate
experienced by user~$k$ under combination~$p$, as well
as~$\text{SR}(n,k,p)$, i.e., the secrecy rate obtained under
permutation~$p$ when the victim is user~$k$ and the eavesdropper is
listening to IRS~$n$. Further, let~$k^\star$ identify the
eavesdropping victim.

Given~$\Pc$, we have to decide which permutations to use, by setting binary variables~$y(p)\in\{0,1\}$; 
also, let~$\bar{\Pc}\subseteq\Pc$ be the set of used permutations. From the decisions~$y(p)$, 
we can write the probability~$\omega(k,n)$ that user~$k$ is served through IRS~$n$ under {\em any} 
of the chosen permutation, i.e.,
\begin{equation}
\omega(k,n)=\frac{\sum_{p\in\bar{\Pc}}\ind{\nu_p(k)=n}}{\left|\bar{\Pc}\right|}.
\end{equation}

\begin{figure*}
\centering
\subfigure[\label{fig:combined-p5}]{
    \includegraphics[width=.32\textwidth]{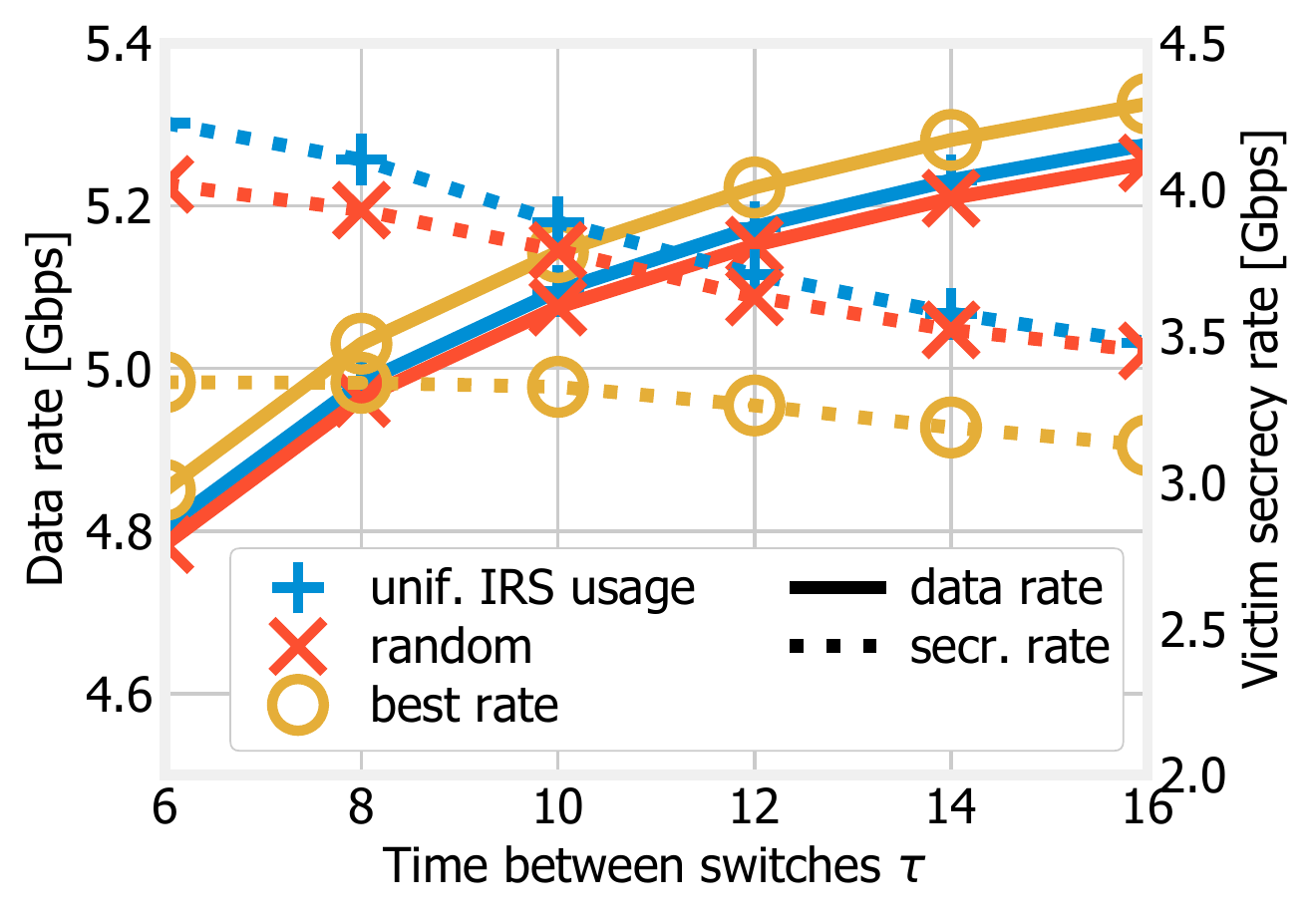}
}
\hspace{-3.5mm}
\subfigure[\label{fig:combined-p10}]{
    \includegraphics[width=.32\textwidth]{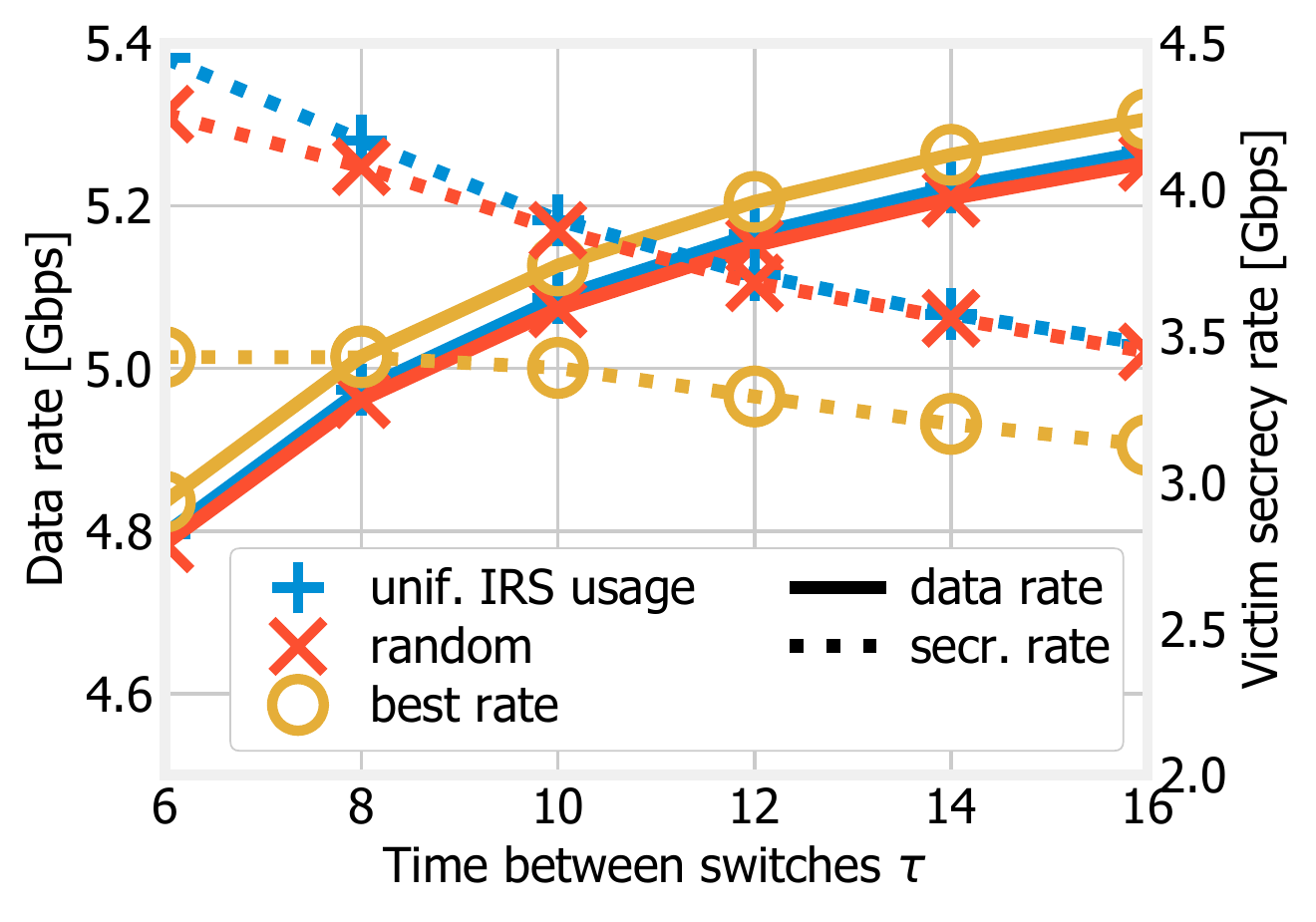}
}
\hspace{-3.5mm}
\subfigure[\label{fig:combined-p20}]{
    \includegraphics[width=.32\textwidth]{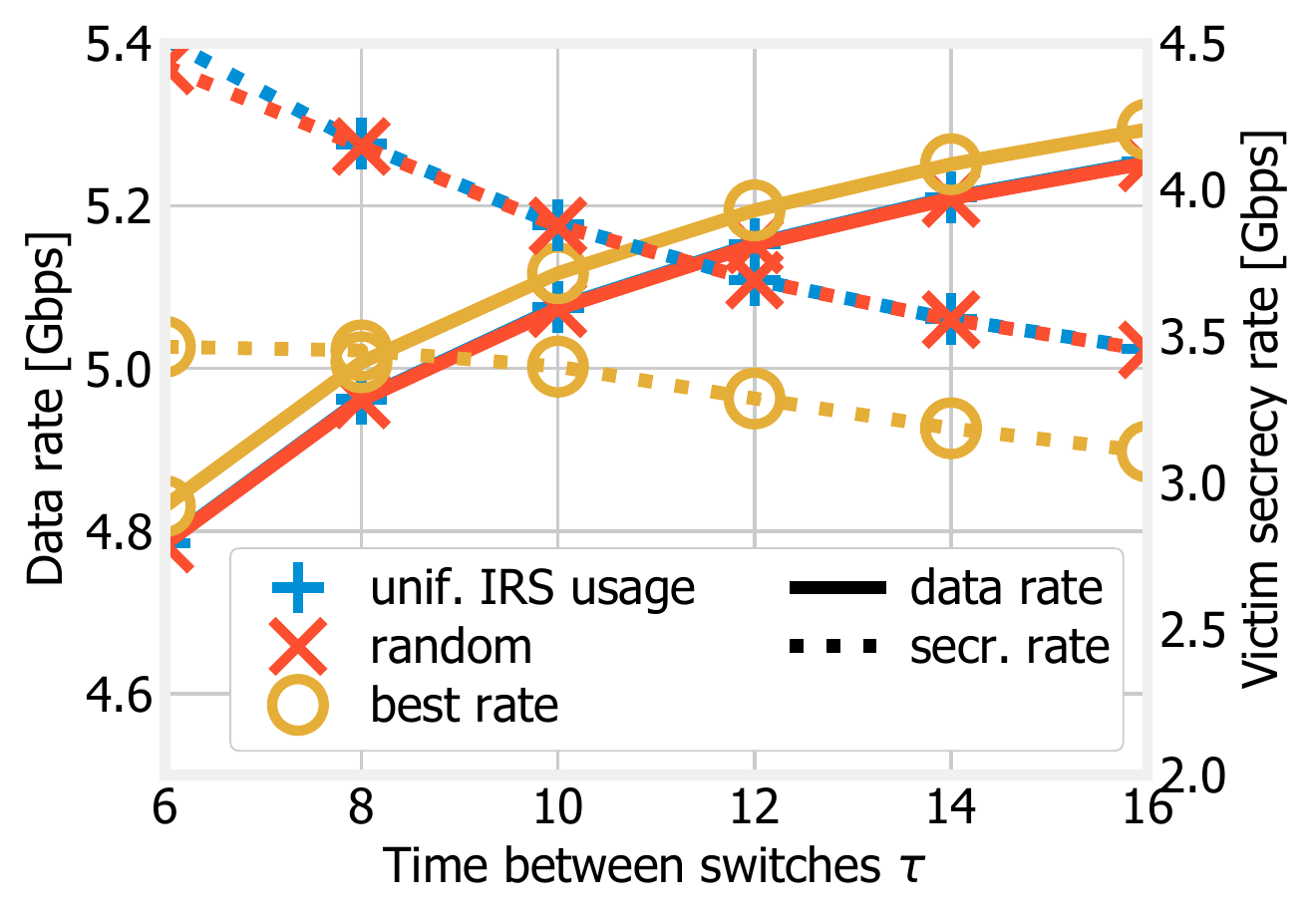}
}
\caption{
Data rate (solid lines) and secrecy rate (dotted lines) as a function of the interval~$\tau$ 
between permutation changes, under different strategies for choosing~$\bar{\Pc}$, 
when the number of active permutations is $|\bar{\Pc}|=5$~(a), $|\bar{\Pc}|=10$~(b), and $|\bar{\Pc}|=20$~(c).
    \label{fig:combined}
}
\end{figure*}

The second decision 
to make is the time~$\tau\geq 1$ for which each permutation is applied
before switching to a new one.  If the time needed for
switching permutation is equal to one time unit and the communication
is paused during such switching time, (i.e., every~$\tau+1$), the {\em
  average} rate for each legitimate user~$k$ is given by:
\begin{equation}
\label{eq:rate-avg}
R_\text{avg}(k)=\frac{\tau}{\tau+1}\frac{\sum_{p\in\bar{\Pc}}R(k,p)}{\left|\bar{\Pc}\right|}\,.
\end{equation}
As expected,  \Eq{rate-avg} tells us that having a small~$\tau$, i.e., switching between permutations too frequently, 
hurts the performance.

Moving to the eavesdropper, its objective is to have the smallest possible secrecy rate (SR) 
for its victim~$k^\star$. There are two strategies it can follow towards this end:
\begin{itemize}
    \item {\em static}: always pointing to the IRS that is most frequently used 
    to serve the victim~$k^\star$, i.e., $n^\star=\arg\max_n\omega(k^\star,n)$, or
    \item {\em dynamic}: ``invest'' $\delta$ time units to try all IRSs, identify the one serving 
    the victim~$k^\star$, and then point towards it.
\end{itemize}
In the first case, the resulting SR is given by:
\begin{equation}
\label{eq:sr-static}
\text{SR}_\text{avg}^\text{static}(k^\star)=\frac{1}{\left|\bar{\Pc}\right|}
\sum_{p\in\bar{\Pc}}\text{SR}(n^\star,k^\star,p)\,,
\end{equation}
while in the latter case, the SR is:
\begin{equation}
\label{eq:sr-dynamic}
\text{SR}_\text{avg}^\text{dynamic}(k^\star)\mathord{=}\frac{1}{\left|\bar{\Pc}\right|}
  \displaystyle\sum_{p\in\bar{\Pc}}\left[\frac{\tau\mathord{-}\delta}{\tau}\min_n\text{SR}(n, k^\star,p)\mathord{+}\frac{\delta R(k^\star,p)}{\tau} \right] 
\end{equation}
if $\delta \le \tau$, and  $\text{SR}_\text{avg}^\text{dynamic}(k^\star) = \frac{1}{\left|\bar{\Pc}\right|}R(k^\star,p)$ otherwise.

The quantity within square brackets in \Eq{sr-dynamic} comes from the
fact that, for each permutation (i.e., each $\tau$~time units), the
eavesdropper spends~$\delta$ units trying all IRSs (during which the
secrecy rate will be~$R(k^\star,p)$, i.e., complete secrecy), and $\tau-\delta$ units experiencing the minimum secrecy rate across
all IRSs.

In both cases, SR values are {\em subordinate to the fact that the
  BS is transmitting} -- clearly, if there is no
transmission, there can be no secrecy rate. Also notice how we must
write SR values as dependent upon the eavesdropping victim~$k^\star$:
indeed, the eavesdropper knows who its victim is, while legitimate
users do not.

The eavesdropper will choose the strategy that best suits it, i.e., results in the lowest secrecy rate. 
It follows that the resulting secrecy rate is:
\begin{equation}
\nonumber
\text{SR}_\text{avg}(k^\star)=\min\left\{\text{SR}_\text{avg}^\text{static}(k^\star),\text{SR}_\text{avg}^\text{dynamic}(k^\star)\right\}.
\end{equation}

The network high-level goal is to maximize the secrecy rate, 
so long as all legitimate users get at least an average rate~$R_{\min}$:
\begin{equation}
\label{eq:obj}
\max_{\bar{\Pc},\tau}\min_{k}\text{SR}_\text{avg}(k),
\end{equation}
\begin{equation}
\label{eq:constr}
\text{s.t.}\quad R_{\text{avg}}(k)\geq R_{\min},\quad\forall k.
\end{equation}
It is interesting to remark how objective \Eq{obj} must be stated in a max-min form: 
indeed, since the BS does not know who the eavesdropping victim is, 
it aims at maximizing the secrecy rate in the worst-case scenario, in which the node with the lowest SR 
is indeed the victim.

Owing to the complexity of choosing the {\em optimal} set~$\bar{\Pc}$ of permutations to use, in the following we will study three heuristics to make such a decision, based upon (i) data rate, (ii) IRS usage, or (iii) random. Indeed, as 
further discussed next, our goal is to investigate the trade-off existing 
between secrecy rate and system throughput when a switching-configuration scheme is adopted 
at the IRS in order to achieve the above objective.

\begin{figure*}
\subfigure[\label{fig:secrate-p5}]{
    \includegraphics[width=.32\textwidth]{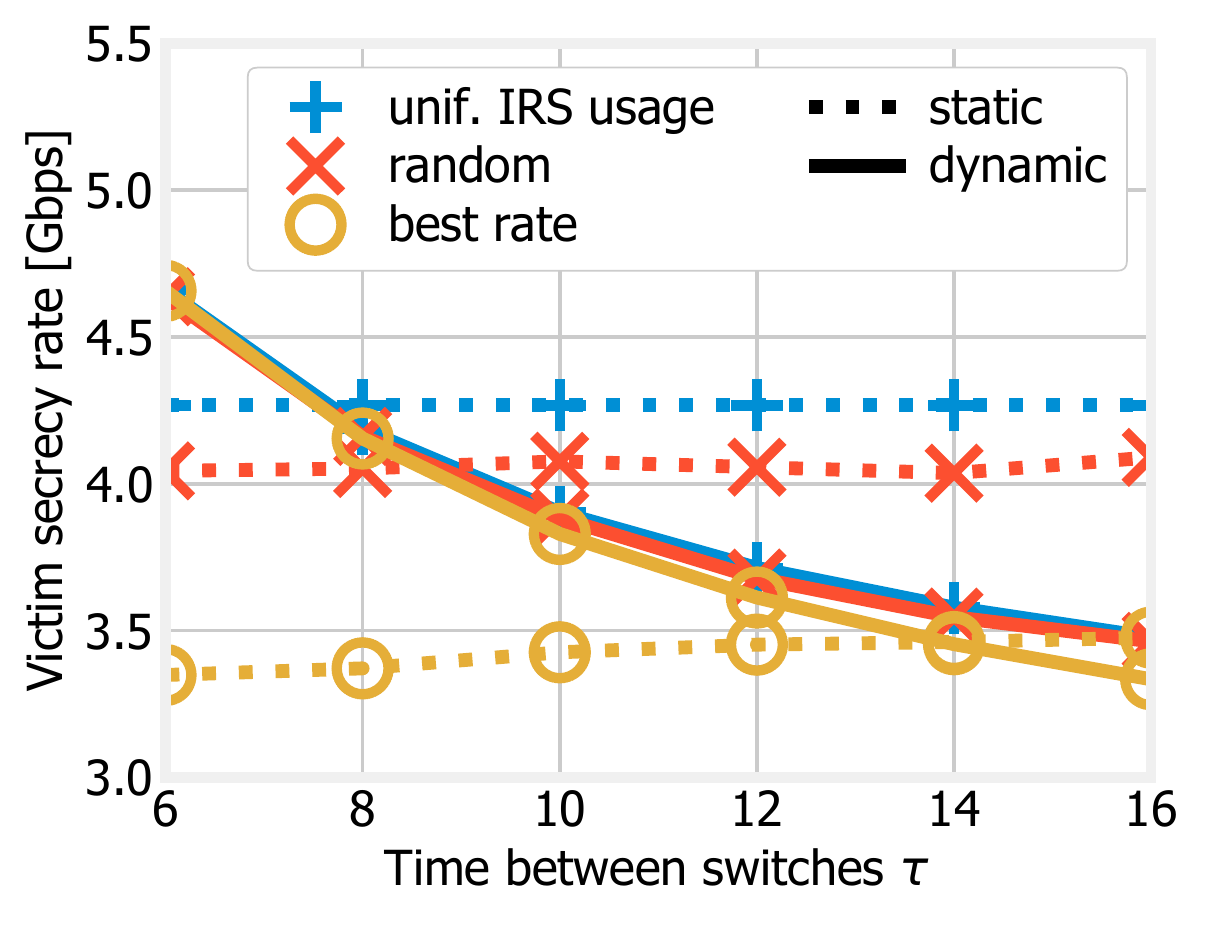}
}
\hspace{-3.5mm}
\subfigure[\label{fig:secrate-p10}]{
    \includegraphics[width=.32\textwidth]{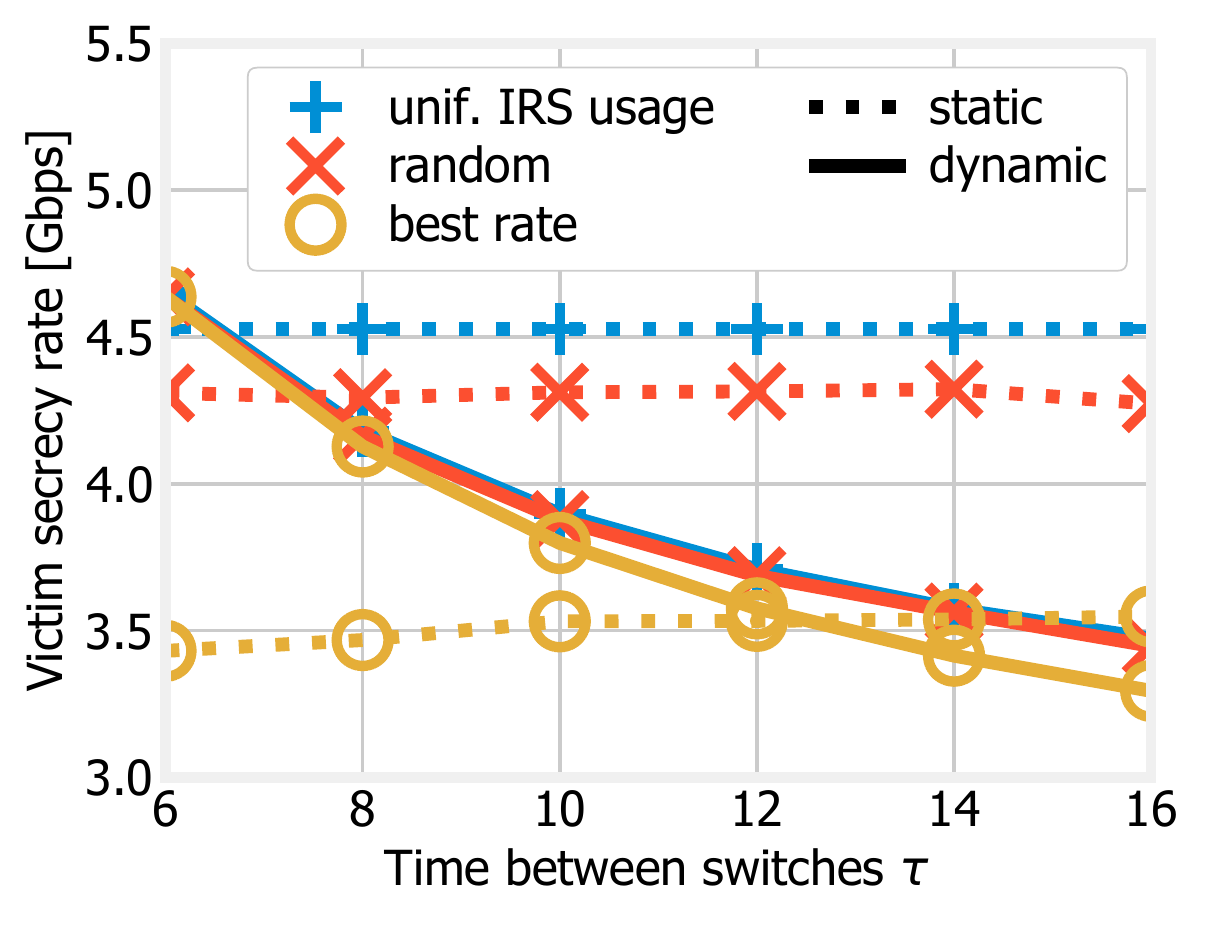}
}
\hspace{-3.5mm}
\subfigure[\label{fig:secrate-p20}]{
    \includegraphics[width=.32\textwidth]{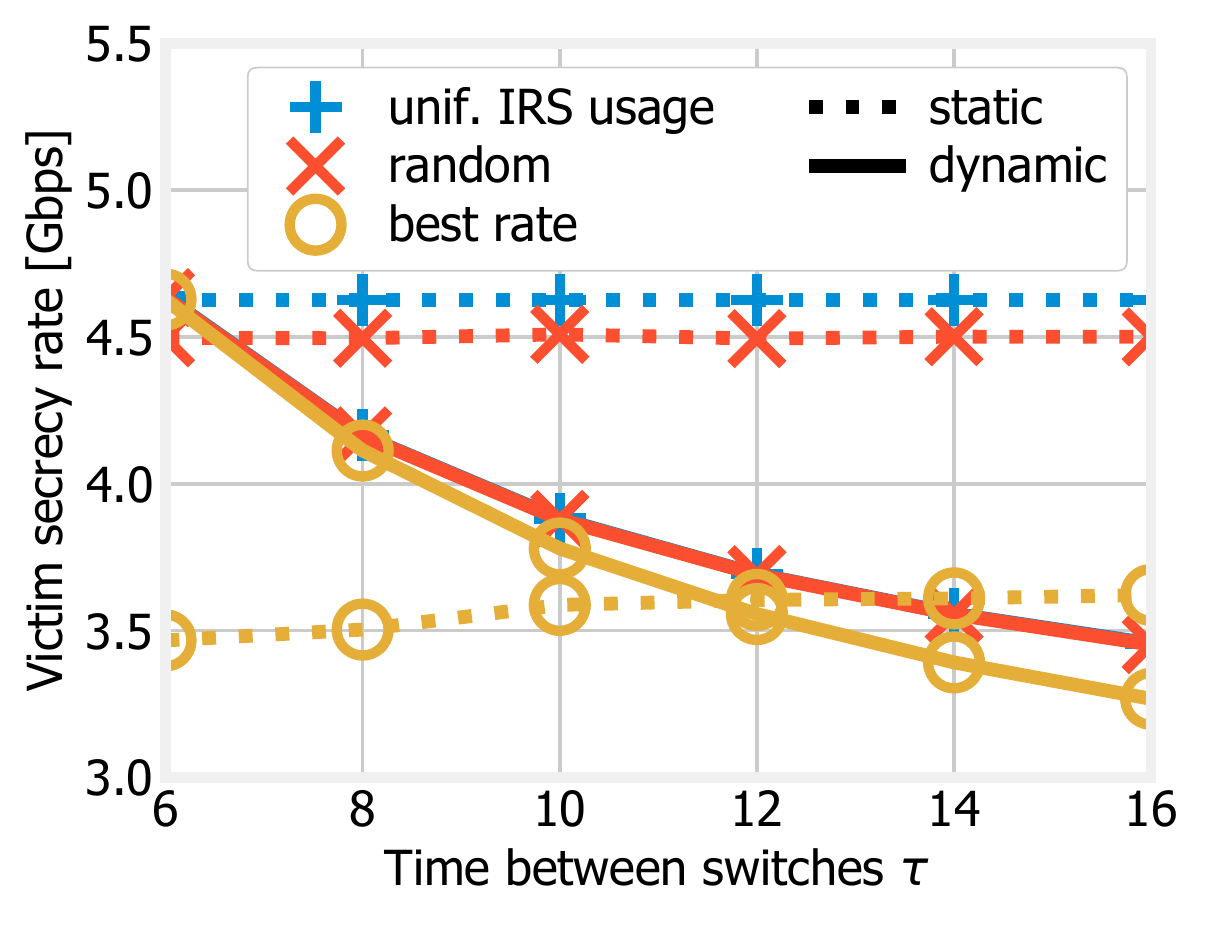}
}
\caption{
Secrecy rate under the static (dotted lines) and dynamic (solid lines) strategies, as a function of the interval~$\tau$ between permutation changes, under different approaches to choose~$\bar{\Pc}$, when the number of active permutations is $|\bar{\Pc}|=5$~(a), $|\bar{\Pc}|=10$~(b), and $|\bar{\Pc}|=20$~(c).
    \label{fig:secrate}
}
\end{figure*}
\begin{figure*}
\subfigure[\label{fig:omega-p5}]{
    \includegraphics[width=.32\textwidth]{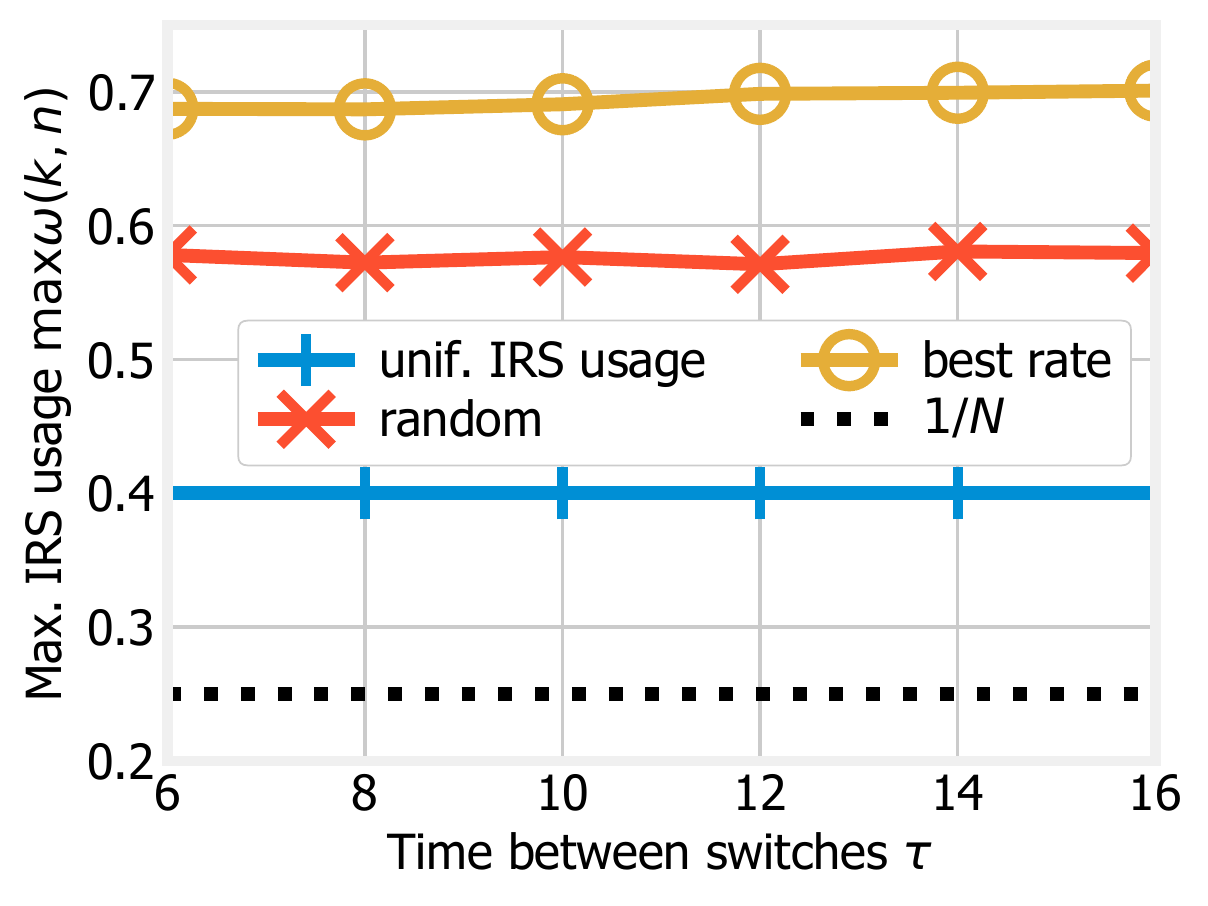}
}
\hspace{-3.5mm}
\subfigure[\label{fig:omega-p10}]{
    \includegraphics[width=.32\textwidth]{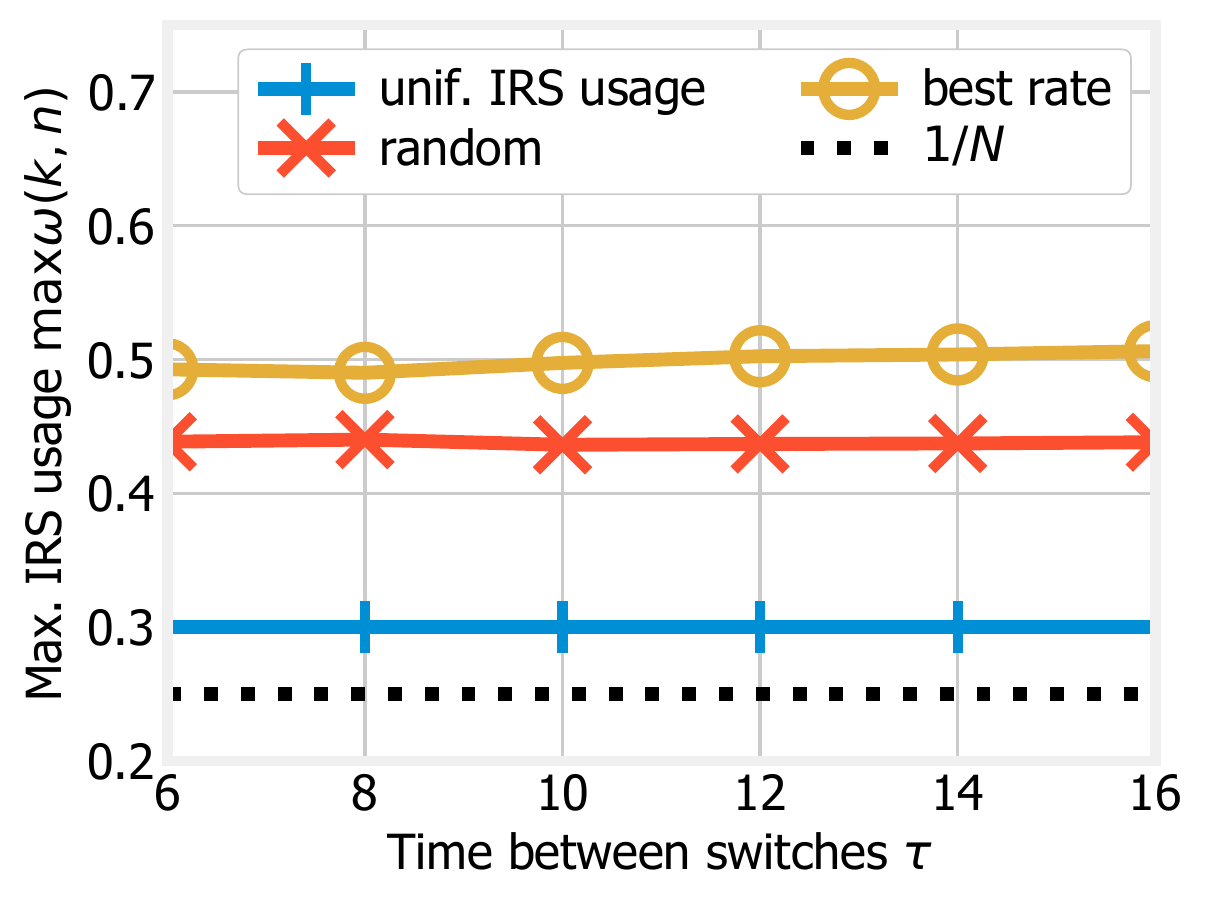}
}
\hspace{-3.5mm}
\subfigure[\label{fig:omega-p20}]{
    \includegraphics[width=.32\textwidth]{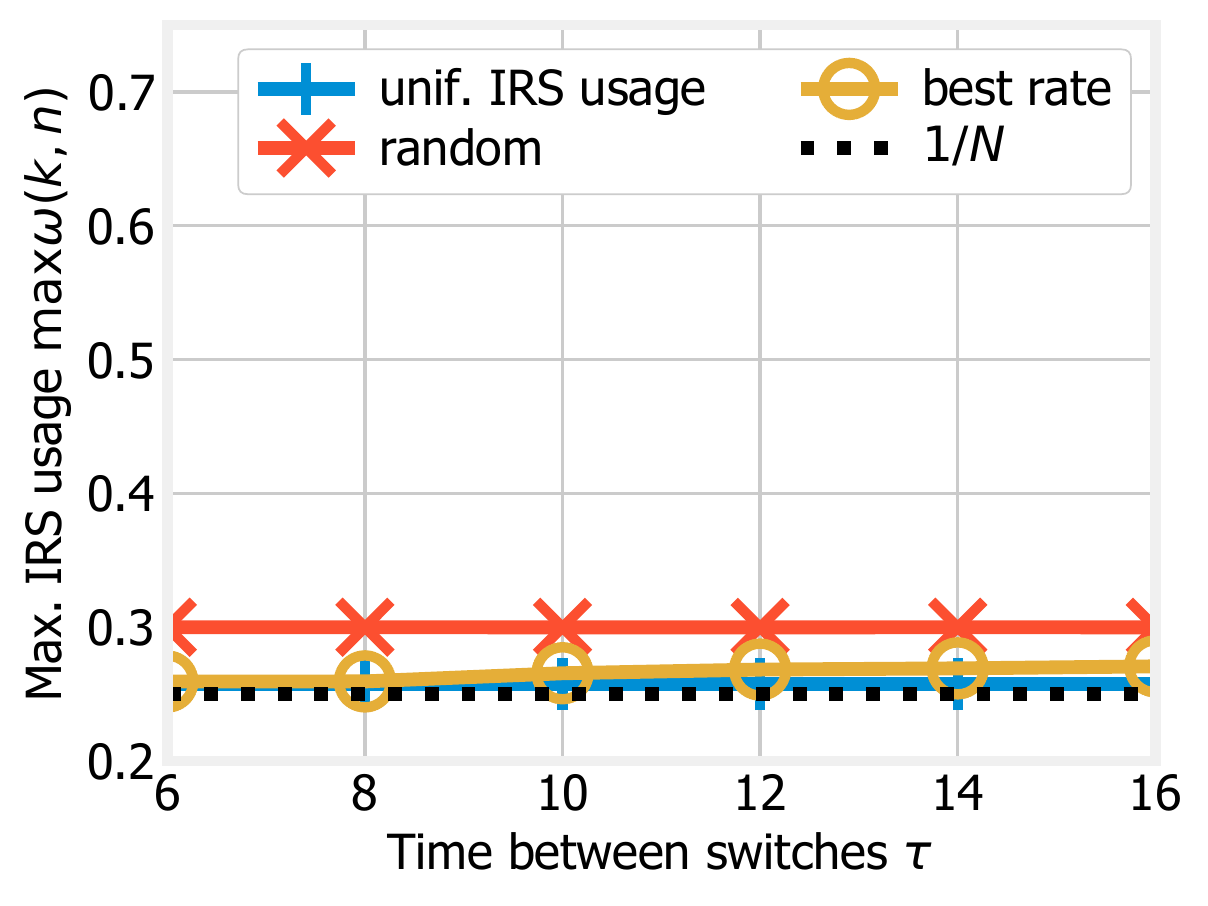}
}
\caption{
Maximum IRS usage, i.e., $\max_{k,n}\omega(k,n)$ as a function of the interval~$\tau$ between permutation changes, under different approaches to choose~$\bar{\Pc}$, when the number of active permutations is $|\bar{\Pc}|=5$~(a), $|\bar{\Pc}|=10$~(b), and $|\bar{\Pc}|=20$~(c).
    \label{fig:omega}
}
\end{figure*}

\section{Performance Evaluation}
\label{sec:peva}

In this section, we seek to characterize and understand the possible
trade-offs between the data rate and secrecy rate, and how the
decisions we make, i.e., the choice of~$\bar{\Pc}$ and~$\tau$,
influence both.

In order to obtain clear, easy-to-interpret results, we focus on a small scale scenario including:
\begin{itemize}
\item one BS, placed at coordinates~$(0,0)$, equipped with
  $M_{\rm BS}=32$ antennas; in our setting the transmit power is
  $P_t=30$\,dBm, the signal wavelength $\lambda=3$\,mm, (corresponding
  to the carrier frequency $f_c=$100\,GHz), the signal bandwidth is
  $B=1$\,GHz, and $N_0=-174$\,dBm/Hz. Moreover, in~\eqref{eq:Gamma} we
  set $\Qm=\Id_K$ so that all UEs experience the same SNR.
    \item $K=4$ UEs, randomly distributed over a $10\times 10$~m$^2$ surface, 
    and equipped with $M_{\rm UE}=4$ antennas;
    \item $N=4$ IRSs of size $64 \times 64$ meta-atoms, randomly placed over the northern wall;
    \item an eavesdropper, randomly placed within 1\,m from its intended victim, 
    which is always UE~$1$ (i.e., $k^\star=1$). The eavesdropper 
    is also equipped with $M_{\rm MN}=4$ antenna elements.
\end{itemize}
As~$N=K=4$, there are~$|\Pc|=4!=24$ possible permutations.
Since we assume that no direct LoS communication is possible between the BS and the UEs, 
each UE is always associated with an IRS that forwards to the UE the signal from the BS.  
Finally we conside~$R_{\min}=0$ and, for simplicity, ignore fading effects.

Throughout our performance evaluation, we compare the following approaches to choose the set~$\bar{\Pc}$:
\begin{itemize}
    \item {\bf best rate}: the $|\bar{\Pc}|$~permutations with the highest data rate are selected;
    \item {\bf uniform IRS usage}: the permutations are chosen so that each UE is served by each IRS 
    with (approximately) the same probability, with ties broken by selecting the highest-rate permutation;
    \item {\bf random}: permutations are chosen at random, with uniform probability.
\end{itemize}

\Fig{combined} shows the data rate and secrecy rate, represented by solid and dotted lines respectively, 
as a function of the interval~$\tau$ between configuration changes, and for different numbers of active permutations, 
i.e., different values of~$|\bar{\Pc}|$. A first aspect we observe is that moving towards larger values of~$|\bar{\Pc}|$, 
i.e., a higher number of active permutations, decreases the data rate while increasing the secrecy rate. 
The reason for the former effect can be seen from the expression of the average rate expression \Eq{rate-avg}, 
which is maximum when only one permutation is active, namely, the one with the highest rate. 
Conversely, having more active permutations makes the work of the malicious node harder, hence, 
it results in a higher secrecy rate.

Furthermore, it is possible to see how the average rate (solid lines in the plots) increases for higher 
values of~$\tau$, i.e., when permutations are kept for a longer time. This is consistent with \Eq{rate-avg}; 
intuitively, if the same network configuration is kept for a longer time, the effect of configuration switches 
is less significant. For similar reasons, the secrecy rate (dotted lines in the plots) decreases  
as $\tau$~increases: once the eavesdropper has identified the best IRS to point to, 
a higher value of~$\tau$ means that it has more time to successfully intercept the communication.

Comparing the different approaches to choose~$\bar{\Pc}$, it is possible to once more observe 
the intrinsic conflict between data rate and secrecy rate. The {\em best rate} approach results, predictably, 
in the highest data rate, but also in a very low secrecy rate. On the other hand, the {\em uniform IRS usage} 
approach yields a very good secrecy rate, at the cost of a data rate which barely exceeds that of {\em random}.

We now focus on the eavesdropper and its behavior. To this end, \Fig{secrate} portrays the secrecy rate 
resulting from the two eavesdropping strategies discussed in \Sec{problem}, namely, static (dotted lines) 
and dynamic (solid lines). It is clear that the dynamic strategy results in a very high secrecy rate when 
$\tau$~is small, i.e., permutations are changed frequently, and becomes more advantageous 
(for the eavesdropper) as $\tau$~increases. 
Such a behavior makes intuitive sense: the dynamic strategy is predicated on investing $N$~time units to find out 
the best IRS to point to, and doing so is pointless if the configuration will change soon afterwards. 
As for the static strategy, the resulting secrecy rate does not depend, as per \Eq{sr-static}, upon~$\tau$, 
hence, its performance is relatively constant.

Interestingly, the value of~$\tau$ for which the curves in \Fig{secrate} overlap, i.e., 
the eavesdropper will move from the static to the dynamic strategy, depends upon the number~$|\bar{\Pc}|$ 
of active permutation, as well as upon the approach followed to choose them. 
This further highlights the nontrivial way in which the decisions of the legitimate and malicious nodes 
interact, even in comparatively simple scenarios like the one we consider.

Last, \Fig{omega} depicts the maximum IRS usage, i.e., the quantity~$\max_{k,n}\omega(k,n)$; 
such a quantity appears in \Eq{sr-static}, hence, it is linked to how strong the static strategy is. 
Intuitively, keeping~$\max_{k,n}\omega(k,n)$ as low as possible makes the static strategy less 
beneficial for the eavesdropper; since, as per \Fig{secrate}, the static strategy is often the most effective, 
doing so has the potential to increase the overall secrecy rate, especially for low values of~$\tau$. 
Indeed, we can see from \Fig{omega} that the {\em uniform IRS usage} strategy results in the smallest 
maximum IRS usage, which correspond to the highest secrecy rates in \Fig{combined} and \Fig{secrate}.

This further confirms the notion emerging from our whole performance evaluation, namely, 
that selecting multiple active permutations and frequently switching between them is necessary 
to thwart the eavesdropper, however, both actions come at a price in terms of network performance, 
i.e., data rate. This, in turn, highlights the need for a deeper understanding of the effects of 
the decisions made by both the BS and the malicious nodes, and for 
algorithms able to leverage such insight.

\section{Conclusions}
\label{sec:concl}
We considered a base station transmitting towards users through the help of intelligent reflecting 
surfaces, and the presence of a passive eavesdropper overhearing the data stream destined to a 
legitimate user. 
Given that the network aims at maximizing the secrecy rate in the system, 
we consider a solution strategy that, leveraging the different possible IRS configurations, 
lets IRS and legitimate users switch from one configuration to another with a given periodicity. 
Importantly, such a scheme (i) exhibits low-complexity, (ii) is suitable for scenarios where 
the nodes have a limited number of antenna elements, and (iii)
provides high secrecy rate at a small cost in terms of 
data rate degradation for legitimate users.
The latter point is demonstrated through a numerical evaluation, which aims at studying the way IRSs are {\em used} and 
at highlighting the trade-off existing between secrecy rate and system throughput.

\bibliographystyle{IEEEtran}
\bibliography{refs}%

\end{document}